\documentclass[a4paper]{jpconf}

\usepackage{amsmath}
\usepackage{amssymb}
\usepackage{graphicx}

\begin{document}

\title{Collision Rate and Symmetry Factor in Gluon Plasma}
\author{Jian Deng}
\address{School of Physics, Shandong University, Shandong 250100, People's Republic
of China}
\author{Qun Wang}
\address{Interdisciplinary Center for Theoretical Study and Department of Modern
Physics, University of Science and Technology of China, Anhui 230026,
People's Republic of China}

\begin{abstract}
The elastic and inelastic collision rates in a gluon gas are calculated. The symmetry factor and the phase space integral are discussed in detail. With a symmetry factor and well constrained phase space, the same result can be obtained as that of the full phase space without the factor.  Such an equivalence is illustrated by analytic and numerical calculations for $gg \rightarrow gg$ and $gg\rightarrow ggg$ processes.      
\end{abstract}

It has been shown in AdS/CFT that the shear viscosity to entropy density ratio $\eta/s$ in a physical system has a universal minimal bound $1/(4\pi)$ \cite{Kovtun:2004de}.
A recent perturbative QCD calculation \cite{Xu:2007ns,Wesp:2011yy} for a gluon plasma shows that such a limit $\eta/s\sim1/4\pi$ 
can be reached by the inelastic process $gg \leftrightarrow ggg$ $(23)$, and the 23 process is 7 times more important than the $gg \leftrightarrow gg$ (22) process. It is in contradiction with the result by Arnold, Moore and Yaffe (AMY) \cite{Arnold:2000dr,Arnold:2003zc}, where the 23 process only provides about a $10\%$ correction to the 22 process. In our previous works \cite{Chen:2009sm, Chen:2010xk, Chen:2011km}, we have pointed out that such a discrepancy might arise from a multiple counting when carrying out collisional integrals of the 23 process over full phase space with a symmetry factor. In this note, we will provide a cross check about the above implication by calculating collision rates with and without symmetry factors.  
We will illustrate the equivalence of two ways of carrying out collisional integrals, in the constrained phase space with the symmetry factor or in full phase space without the symmetry factor.

For a hot gluon plasma in the vicinity of thermal equilibrium, the collision rates are defined by the detailed balance rate in Boltzmann equation \cite{Chen:2009sm}. For the 22 process, due to the symmetry between the $u$-channel and $t$-channel in the matrix element $|M_{12\rightarrow34}|^2$, we may choose one channel in the the phase space integration and multiply the result by a symmetry factor of 2. For example, if we choose the $t$-channel and limit the phase space to the near forward angle scatterings, where $\mathbf{q}=\mathbf{p}_3-\mathbf{p}_1$, and $\mathbf{q}_T$ is the transverse component of $\mathbf{q}$ with respect to $\mathbf{p}_1$ in the CM frame, the matrix element reads  
\begin{eqnarray}
|M_{12\rightarrow34}|^2_{CM} \underset{\mathbf{q}^{2}\approx \mathbf{q}_{T}^{2}\approx 0}{\approx } 144 \pi^2 \alpha_s^2 
\frac{s^2}{(\mathbf{q}_T^2+m_D^2)^2}.\label{ME22-1}
\end{eqnarray}
Here $\alpha_s=g^2/(4\pi)$ is the strong coupling constant, and we have inserted the Debye mass $m_D\sim gT$ to regularize gluon propagator. If we do not limit the phase space in the integration, i.e. we take full phase space including both channels, 
then the factor of 2 is not needed:   
\begin{equation}
|M_{12\rightarrow 34}|_{CM}^{2}\underset{\mathbf{q}_{T}^{2}\approx 0}{
\approx }72 \pi^2 \alpha_s^2\frac{s^{2}}{(\mathbf{q}
_{T}^{2}+m_{D}^{2})^{2}}.   \label{ME22-2}
\end{equation}
Note that the constraint $\mathbf{q}^{2}\approx 0$ is removed because both
the near forward and backward scatterings have small $\mathbf{q}_{T}^{2}$
but only the near forward scatterings have small $\mathbf{q}^{2}$ \cite{Chen:2011km}.

\begin{figure}[tbp]
\includegraphics[scale=0.6]{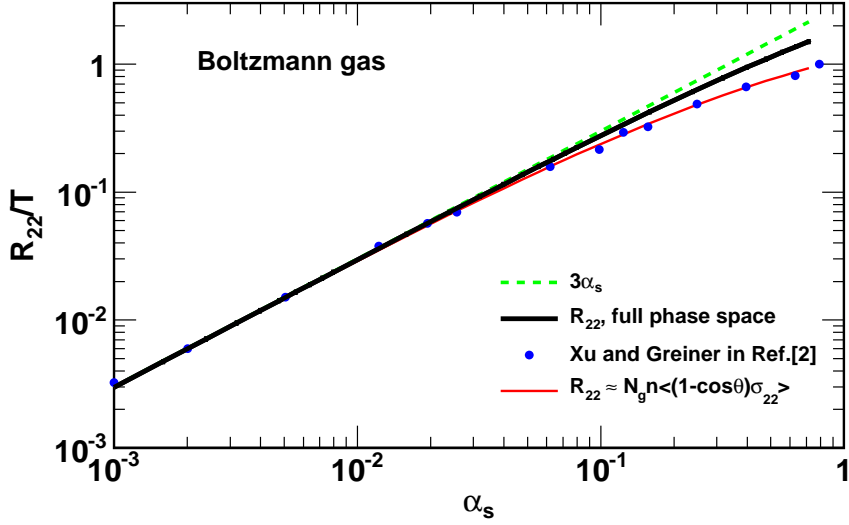}
\caption{The elastic collision rate $R_{22}$ in a Boltzmann gas.}\label{fig:1}
\end{figure}

With the matrix element, it is straightforward to calculate the elastic collision rate. Using the small $\mathbf{q}$ approximation, the integral of $d^3p_3 d^3p_4$ can be simplified as
\begin{eqnarray} \label{delta22}
\int \frac{d^3p_3}{(2\pi)^3}\frac{d^3p_4}{(2\pi)^3} (2\pi)^4 \delta^4(p_1+p_2-p_3-p_4)[\cdots] \approx \frac{1}{4\pi^2}\int d^3 q \delta(\hat{\mathbf{k}_1}\cdot \mathbf{q}-\hat{\mathbf{k}}_2\cdot \mathbf{q})[\cdots].
\end{eqnarray} 
The $\delta$ function can be removed by the integral over $\mathbf{q}$, further integral over $\mathbf{p}_1$ and $\mathbf{p}_2$ gives the result $R_{22}=\frac{\pi^2}{2\zeta(3)}\alpha_s T \approx 4.105\alpha_s T$ for a boson gas with $m_D^2=4\pi \alpha_s T^2$. For a Boltzmann gas with $m_D^2=24/\pi \alpha_s T^2$, the Bose enhancement factor $(1+f_3^0)(1+f_4^0)$ is absent, and the result is  $R_{22}=3\alpha_s T$. To check the above analytic results, we can carry out the collisional integral numerically using the matrix element in Eq.(\ref{ME22-2}) in full phase space. The result is shown in Fig.\ref{fig:1} by the thick solid line, which matches the dashed line of $R_{22}=3\alpha_s T$ at $\alpha_s <0.02$ perfectly. The difference between the two curves for large $\alpha_s$ is due to the fact that the small $\mathbf{q}$ approximation doesn't hold for large $m_D$.         

Another way of calculating the rate is through the cross section \cite{Xu:2007aa} with which the elastic collision rate in a Boltzmann gas can be written as, 
\begin{eqnarray}
R_{22}=\frac{N_g}{n}\int \frac{d^3p_1}{(2\pi)^3}\frac{d^3p_1}{(2\pi)^3}\frac{p_1\cdot p_2}{E_1 E_2}f_1^0 f_2^0\sigma_{22}=N_g n<(1-\cos\theta_{12})\sigma_{22}>.
\end{eqnarray}
If neglecting the $s$ dependence in $\sigma _{22}$, it is easy to get $R_{22}=3\alpha_s T$. With the $s$ dependence, the average over the $\mathbf{p}_1$ and $\mathbf{p}_2$  gives the thin solid line in Fig.\ref{fig:1}, which agrees with the Xu and Greiner's result \cite{Xu:2007ns} quite well.

\begin{figure}[tbp]
\includegraphics[scale=0.6]{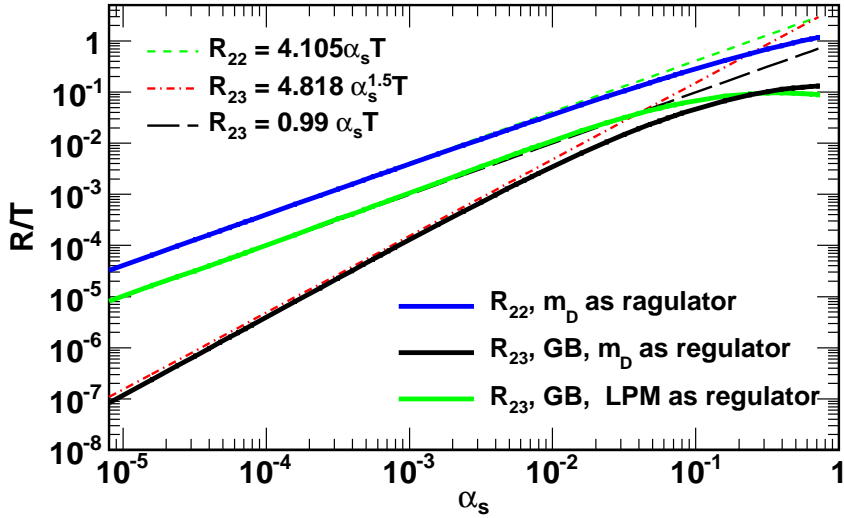}
\caption{The elastic and inelastic collision rate for a boson gas. Solid lines stand for the numerical results in full phase space without the symmetry factor, long-dashed, short-dashed and dashed-dotted line stand for semi-analytic results in the constrained phase space with the symmetry factor.}\label{fig:2}
\end{figure}

For the 23 process, the exact tree level matrix element \cite{Berends:1981rb,Ellis:1985er,Gottschalk:1979wq} is given, e.g., in Eqs. (5,6,12) of Ref. \cite{Chen:2011km}. In the limit of one particular configuration in phase space, namely, $p_1,p_2,p_3,p_4$ are hard (their momenta are $O(\sqrt{s})$), while $q(=p_2-p_4=p_3+p_5-p_1)$ and  $k=p_5$ are soft (momenta are much smaller than $\sqrt{s}$), the matrix element becomes   
\begin{equation}
\left\vert M_{12\rightarrow 345}\right\vert _{CM}^{2}\approx 3456 \pi^3 \alpha_s^3\frac{s^{2}}{\left( k_{T}^{2}+m_{g }^{2}\right)
\left( q_{T}^{2}+m_{D}^{2}\right) \left[ \left( \mathbf{k}_{T}-\mathbf{q}
_{T}\right) ^{2}+m_{D}^{2}\right] }.  \label{reduce-GB}
\end{equation}
It can be seen that the Gunion-Bertsch (GB) formula \cite{Gunion:1981qs} is recovered after taking $m_D, m_g \rightarrow 0$, and in this limit, the denominator can be written in a symmetric form $\mathbf{p}_{T3}^2 \mathbf{p}_{T4}^2 \mathbf{p}_{T5}^2$. There are 6 different configurations in the full phase space, represented by the permutation of $\{3,4,5\}$. If the phase space of a single configuration can be isolated and integrated in collisional integrals, the result times 6 should be equal to the full phase space result. We will use the above configuration where $k_5$ is soft  
to calculate the rate of the 23 process with the symmetry factor. We now take some approximations as follows: (1) For small $k$, the Bose enhancement diverges as $1+f_k \sim 1/k$. (2) As the contribution from small $k$ is enhanced by the divergence, we may assume $k\ll q$ and $|\mathbf{k}_T| \ll |\mathbf{q}_T|$, then the integrals over $k$ and $q$ are decoupled.  (3) The recoil from soft gluon to hard gluons is negligible, we have $\delta^4(p_1+p_2-p_3-p_4-p_5)\approx \delta^4(p_1+p_2-p_3-p_4)$, and Eq.(\ref{delta22}) is still valid. So the inelastic collision rate can be written as           
\begin{eqnarray}  \label{semiR23-1}
R_{23}&=&  6 \times \frac{N_g 3456 \pi^3 \alpha_s^3}{6 n} \int p_1^2 dp_1 (2\pi) \sin(\theta_{12})d\theta_{12}\int p_2^2 dp_2 (4\pi) \nonumber\\
&&\times\int k^2 dk \sin(\theta_k)d\theta_k d\phi_k \int q^2 dq \sin(\theta_q)d\theta_q d\phi_q \nonumber \\
&&\times \frac{\delta(\phi_q-\theta_{12}/2)}{q\sin{\theta_q}\sin(\theta_{12}/2)}\frac{f_1 f_2 (1+f_1) (1+f_2) (1+f_k)}{(2\pi)^{11}2^5E_1^2 E_2^2 k}\nonumber\\
&&\times\frac{s^2}{(\mathbf{q}_T^2+m_D^2)^2}\cdot \frac{\mathbf
{q}_T^2}{(\mathbf{k}_T^2+{m_D^2})(\mathbf{q_T}^2+m_D^2)}. 
\end{eqnarray}
Note that symmetry factor 6 is added, and the GB formula for the matrix element in literature is used. Although complicated enough, most of the integral can be integrated out. We will end up with three entangled angular integrals which have to be computed numerically. The final result gives $R_{23}=4.818 \alpha_s^{3/2}$. Note that the divergence of Bose distribution in soft regime plays a very important role. In a Boltzmann gas, these approximations do not hold and the analytical approach do not work. In a general case, we should integrate over full phase space without the symmetry factor by solving $\delta^4(p_1+p_2-p_3-p_4-p_5)$ exactly and numerically. The results from two methods are shown as the black solid line and red dashed-dotted line in Fig.\ref{fig:2}. The sizable difference at large $\alpha_s$ comes from the fact that the small $\mathbf{q}_T$ approximation does not hold for large $m_D$.

Another way to regulate the IR singularity is to use the Landau-Pomeranchuk-Migdal (LPM) effect \cite{Gyulassy:1991xb, Wang:1994fx} to provide a cutoff for the bremsstrahlung gluon. One can simply use $\Theta(\mathbf{k}_T^2- E_k/l_{mfp})$ in the collisional integral, where the the mean free path can be estimated by the inverse of the collision rate $l_{mfp}\sim 1/R =1/(R_{22}+2.5 R_{23})$. The semi-analytic method is still workable and gives the self-consistent equation $R_{23}^{LPM}=6.52\alpha_s^2/(R_{22}+2.5R_{23}^{LPM})$, whose solution is $R_{23}^{LPM}=\frac{1}{5}(\sqrt{R_{22}^2+65.2 \alpha_s^2}-R_{22}) \approx 0.99 \alpha_s T$. We can also get a self-consistent result by iterating the numerical integration over the full phase space, in which the LPM condition applied to the softest gluon for each configuration satisfying the energy-momentum conservation. The results from the two methods agree with each other perfectly at small $\alpha_s$ as shown by the green solid and black long-dashed line in Fig.\ref{fig:2}. 

In summary, we have calculated elastic and inelastic collision rates in a gluon gas with different methods to show the significance of the symmetry factor in phase space integrals. The current study confirms our previous result about the role of the symmetry factor in the shear and bulk viscosity \cite{Chen:2009sm, Chen:2010xk, Chen:2011km}. 
We have illustrated the equivalence of carrying out collision integral in the constrained phase space with the symmetry factor or in full phase space without the symmetry factor. Our result shows that the 22 process is more important than 23 process with reasonable coupling constant where the perturbative QCD is reliable. With larger coupling constant, for example, in the QGP created at the RHIC/LHC energies, $\alpha_s\sim 0.2-0.3$, non-perturbative effects should be taken into account to explain the observed small shear viscosity $\eta/s\sim 1/(4\pi)$. The current scheme can be applied to calculate some other thansport coefficients, such as the dragging coefficient of heavy quarks, the jet quenching parameter, etc., on which we will look at in the future. 

\vspace{0.3cm}

QW is supported in part by the National Natural Science Foundation
of China under grant 11125524. JD is supported by the National Natural Science Foundation
of China under grant 11105082.

\end{document}